\documentclass[12pt,preprint]{aastex}

\slugcomment{Accepted for publication in ApJL}

\shorttitle{The XUV disk of NGC~4625}
\shortauthors{Gil de Paz et al.}

\begin{document}

\title{Discovery of an extended UV disk in the nearby galaxy NGC~4625}

\author{
A. Gil de Paz\altaffilmark{1},
B. F. Madore\altaffilmark{1},
S. Boissier\altaffilmark{1},
R. Swaters\altaffilmark{2},
C. C. Popescu\altaffilmark{3}, R. J. Tuffs\altaffilmark{3},
K. Sheth\altaffilmark{4}, R. C. Kennicutt\altaffilmark{5},
L. Bianchi\altaffilmark{6}, D. Thilker\altaffilmark{6}, 
D. C. Martin\altaffilmark{7}
}

\altaffiltext{1}{Observatories of the Carnegie Institution of Washington, 813 Santa Barbara St., Pasadena, CA 91101}
\altaffiltext{2}{Department of Astronomy, University of Maryland, College Park, MD 20742-2421}
\altaffiltext{3}{Max Planck Institut fur Kernphysik, Saupfercheckweg 1, 69117 Heidelberg, Germany}
\altaffiltext{4}{Spitzer Science Center, California Institute of Technology, MC 220-6, 1200 East California Boulevard, Pasadena, CA 91125}
\altaffiltext{5}{Steward Observatory, University of Arizona, 933 North Cherry Avenue, Tucson, AZ 85721}
\altaffiltext{6}{Center for Astrophysical Sciences, The Johns Hopkins University, 3400 N. Charles St., Baltimore, MD 21218}
\altaffiltext{7}{California Institute of Technology, MC 405-47, 1200 East California Boulevard, Pasadena, CA 91125}

\begin{abstract}
Recent far-UV (FUV) and near-UV (NUV) observations of the nearby
galaxy NGC~4625 made by the {\it Galaxy Evolution Explorer} (GALEX)
show the presence of an extended UV disk reaching four times the
optical radius of the galaxy. The UV-to-optical colors suggest that
the bulk of the stars in the disk of NGC~4625 are currently being
formed, providing a unique opportunity to study today the physics of
star formation under conditions similar to those when the normal disks
of spiral galaxies like the Milky Way first formed. In the case of
NGC~4625, the star formation in the extended disk is likely to be
triggered by interaction with NGC~4618 and possibly also with the
newly-discovered galaxy NGC~4625A. The positions of the FUV complexes
in the extended disk coincide with peaks in the H{\sc i}
distribution. The masses of these complexes are in the range
10$^{~3-4}$\,M$_{\odot}$ with their H$\alpha$ emission (when present)
being dominated by ionization from single stars.
\end{abstract}

\keywords{galaxies: formation, star clusters, individual (NGC~4625)---ultraviolet: galaxies}

\section{Introduction}
\label{intro}

The disks of spiral galaxies are thought to have formed stars more or
less continuously over the last 5-10\,Gyr, with their individual
star-formation histories being a function of the mass and angular
momentum of the disk (Bell \& de Jong 2000; Boissier et al$.$
2001). The current specific star formation rates in the disks of most
spiral galaxies are as low as 0.1\,Gyr$^{-1}$ (Boissier et al$.$
2001). Therefore, the only way to learn about the properties of early
spiral galaxies, including the physics of their star formation, was
thought to be by looking at faint, poorly spatially resolved,
high-redshift galaxies.

In this Letter we report the discovery of an extended UV (XUV) disk in
the nearby galaxy NGC~4625 reaching four times its optical radius at
$\mu_{B}$=25\,mag\,arcsec$^{-2}$ (D25 radius hereafter). Although
Swaters \& Balcells (2002) already showed the presence of an extended
low-surface-brightness component in their optical surface-brightness
profiles of this galaxy, the GALEX observations show that the disk is
even more extended at UV wavelengths, that it has a well-defined
spiral morphology (so it is certainly a disk not a halo), and that it
seems to be forming most of its stars at the current epoch. Therefore,
NGC~4625 may provide a rare opportunity to study locally the
conditions governing the early formation of the disks of spiral
galaxies. This object increases the small number of XUV disks reported
to date and constitutes its most extreme example (Thilker et al$.$
2005a, 2005b, in prep.).

NGC~4625 is a nearby (9.5\,Mpc for H$_0$=70\,km\,s$^{-1}$\,Mpc$^{-1}$;
Kennicutt et al$.$ 2003), low-luminosity (M$_B$$\sim$$-$17.4)
one-armed Magellanic spiral galaxy thought to be interacting with the
also single-armed spiral NGC~4618\footnote{Note that Odewahn (1991)
estimated a rather closer distance for this system of 6.0\,Mpc}. In
spite of the interaction with NGC~4618, NGC~4625 shows a remarkably
regular H{\sc i} velocity field and a well-defined rotation curve
(Bush \& Wilcots 2004). To the South-East we report the discovery of a
low-surface brightness galaxy only $\sim$4\,arcmin away from NGC~4625,
which, if its association with NGC~4625 is confirmed, might also have
played a role (along with NGC~4618) in the recent activation of the
star formation in the XUV disk of NGC~4625.

\section{Observations and analysis}
\label{observations}

The NGC~4618/4625 system was simultaneously observed in the FUV and
NUV bands by GALEX on April 5th 2004 for 3268 seconds split across two
orbits. Images were reduced using the GALEX pipeline (Martin et al$.$
2005). The spatial resolution achieved was FWHM$\simeq$4.5 and 
5.0\,arcsec in the FUV and NUV channels respectively.

Deep ground-based optical imaging data in $B$ and $R$ bands were
obtained previously with the prime-focus camera of the 2.5-m Isaac
Newton Telescope (La Palma, Spain) on May 28th 1995 (see Swaters
\& Balcells 2002 for details). On August 20th 2004, we obtained an 
H$\alpha$ image at the Palomar Observatory 5-m telescope using COSMIC
and a 6563/20 narrow-band filter. The total exposure time was 800
seconds. The seeing on all three optical images is in the range
1.2-1.4\,arcsec. Neutral hydrogen observations in the line of 21\,cm
of the NGC~4618/4625 system were obtained with the Westerbork
Synthesis Radio Telescope (WSRT) as part of the WHISP survey
(Kamphuis, Sijbring, \& van Albada 1996; Swaters et al$.$ 2002).

We have derived surface photometry for NGC~4625 in the FUV, NUV, $B$,
and $R$ bands, H$\alpha$, and H{\sc i} using circular annuli centered
on RA(J2000)=12$^{\mathrm{h}}$41$^{\mathrm{m}}$52.6$^{\mathrm{s}}$ and
DEC(J2000)=$+$41$^{\circ}$16'21.5''. Circular apertures provide a very
clear separation between the optical disk (with approximately circular
isophotes) and the XUV disk.  Almost identical results (both
qualitatively and quantitatively) were obtained when elliptical
apertures matching the light distribution of the XUV disk were
used. Individual-region photometry was computed using elliptical
apertures as defined in the FUV image by Sextractor (Bertin
\& Arnouts 1996). A few FUV sources missed by Sextractor were added by
hand to the source catalog.

\section{Discussion}
\label{discussion}

\subsection{XUV-disk morphology and stellar populations}

Figure~1 shows that the XUV emission in NGC~4625 (almost invisible in
shallow optical images from the ground) covers a significant fraction
of the area detected in 21\,cm with some correspondence between the
position of the brightest UV complexes and peaks in the neutral-gas
distribution. The XUV disk is made up of several fragmented spiral
arms in the inner regions and possibly a large faint arm in the
outermost regions (Figure~2a). Deep ground-based images show a similar
morphology at optical wavelengths and also reveal the presence of a
newly-discovered faint, red, low-surface brightness companion [to be
called hereafter NGC~4625A,
RA(J2000)=12$^{\mathrm{h}}$42$^{\mathrm{m}}$11.1$^{\mathrm{s}}$;
DEC(J2000)=$+$41$^{\circ}$15'10''] seen $\sim$4\,arcmin to the
south-east of NGC~4625 (see Figure~2b). Deep optical spectroscopy
would be required to establish if NGC~4625A is physically associated
with NGC~4625; chance projection of such a rare type of object is,
however, unlikely.

The surface photometry of NGC~4625 (Figure~3) shows very blue UV
colors for the innermost regions of the galaxy
[(FUV$-$NUV)$\simeq$0.5\,mag, typical of Magellanic spirals like
NGC~4625; Bell et al$.$ 2002] followed by a distinctly redder zone
falling in the annulus 25-50\,arcsec in radius and coincident with a
steep exponential decline in surface brightness. Dust is not likely to
be responsible for this reddening because (1) the dust content in
NGC~4625 is relatively low overall
[E($B-V$)=0.1\,mag]\footnote{Derived from the global infrared-to-UV
ratio [$\log$(TIR/FUV)=0.23] assuming the relationship between
(TIR/FUV) and A$_{\mathrm{FUV}}$ given by Buat et al$.$ (2005) and a
Galactic extinction law. If the effects of scattering are considered
E($B-V$) could be as low as 0.04\,mag (Tuffs et al$.$ 2004).}, and (2)
the radial profiles of reddening-free color indices, like the
(FUV$-$NUV) color (Bianchi et al$.$ 2005), show a clear change in
color at the same position as well.  If we now consider the fact that
the profile at this position is very smooth and it is not associated
with any H{\sc ii} region or bright UV cluster we then identify this
transition region with an underlying, evolved (intrinsically red)
Population II component\footnote{A similar smooth red envelope is also
seen in the outer parts of NGC~4618 (see Figure~1a)}. The sharply
declining surface brightness profile of NGC~4625 in this region
(between 25-50\,arcsec) suggests that the (extrapolated) contribution
of this population to the even more distant, extended UV emission is
negligible. Finally, we note that the XUV disk shows very blue colors,
especially in (FUV$-$NUV) and (NUV-$B$), with a rather flat profile in
(FUV$-$NUV) and ($B-R$), but a clear bluing in the (NUV$-$$B$)
profile. These colors suggest the presence of a young stellar
population ($<$1\,Gyr) dominating the UV and optical emission (and
probably also the mass) of the XUV disk of NGC~4625. However, although
it seems unlikely (based on the light distribution of its Population
II component) we cannot rule out the presence of a faint
several-Gyr-old stellar population partly contributing to the observed
colors and stellar mass of the XUV disk. Deep near-infrared photometry
from the ground or a single-star-photometry color-magnitude diagram
with {\it Hubble Space Telescope} should provide fundamental clues for
answering this question.

Compared with the distribution of the UV emission, the
azimuthally-averaged H$\alpha$ emission in the XUV disk is much
fainter than that of the innermost regions of the galaxy (see
Figures~2b \& 3). On the other hand, the H{\sc i} emission is clearly
more extended still than the UV. Beam smearing is unlikely to be
responsible for the apparent flattening of the neutral-gas profile
considering the relatively good spatial resolution of our H{\sc i} map
(FWHM$\simeq$15\,arcsec).

A more detailed analysis of the star formation history of NGC~4625 and
of the implications of these results on the star formation law will be
presented in a forthcoming paper (Gil de Paz et al$.$ 2005, in prep.).

\subsection{Stellar complexes}

Inspection of our deep, arcsec-resolution, optical images of NGC~4625
shows that the FUV-bright complexes in the XUV disk are made up of one
or several stellar clusters in which some bright individual stars can
be resolved (Sandage \& Bedke 1994). Aperture photometry on a sample
of 74 FUV-selected complexes in the XUV disk yields UV luminosities in
the range 10$^{~23.0-24.5}$\,erg\,s$^{-1}$\,Hz$^{-1}$ (dots and
diamonds in Figure~4). If we assume that their stars formed
instantaneously, the corresponding stellar masses would be in the
range 10$^{~3-4}$\,M$_{\odot}$. Figures~2b \& 4 also show that most of
the H$\alpha$ emission in the XUV disk comes from very compact sources
with $L_{\mathrm{H}\alpha}$$<$10$^{~37.6}$\,erg\,s$^{-1}$. These
H$\alpha$ luminosities are compatible with being produced by single
stars having masses less than 60\,M$_{\odot}$, or associations of a
few less massive O-stars.

One of the most puzzling properties of the XUV disks recently
discovered by GALEX is the fact that even these galaxies seem to have
a truncation radius in the distribution of the inner-disk H$\alpha$
emission (e.g$.$ Thilker et al$.$ 2005a). Such truncation has been
traditionally explained by a star-formation threshold (Martin \&
Kennicutt 2001). While in NGC~4625 there is no sharp truncation (since
faint H$\alpha$ emission is detected very far out in the disk), the
H$\alpha$ emission does decline with radius much faster than the UV
light (Figure~3). In Figure~4 we compare the H$\alpha$ and FUV
luminosities of individual complexes in the optically-bright disk of
NGC~4625 (inside D25; triangles), in the inner regions of the XUV disk
(between D25 and 130\,arcsec in radius; dots), and in the outer XUV
disk (beyond 130\,arcsec; diamonds). This figure shows that while
regions inside D25 have H$\alpha$ to UV ratios similar to those
expected for continuous star formation models, those in the XUV disk
have somewhat lower H$\alpha$ to UV ratios, especially in the very
outer parts of the XUV disk. FUV-selected complexes in the outer XUV
disk (diamonds) are also fainter in the UV than those in the inner XUV
disk (dots).

We note that the H$\alpha$ to UV ratios derived for individual
complexes in the XUV disk are systematically higher than those found
for the azimuthally-average surface brightness profiles (adopting an
average 7\,arcsec-radius aperture; thick solid magenta line). This can
be explained by the presence of diffuse UV emission from stars of
intermediate age (a few 10$^{8}$\,yr) in the inter-arm region or from
scattering by dust grains (Popescu et al$.$ 2005). If the former case
is true it would imply that the star formation in the XUV disk is
triggered by gravitational instabilities in the form of low-order
spiral density waves or tidal interactions, which have time-scales of
the order of the UV emission (and also comparable to the galaxy
dynamical time-scale) but much longer than that of H$\alpha$. In the
central parts of the optical disk, on the other hand, the higher
frequency and efficiency of the gravitational instabilities and the
presence of sequential and turbulence triggering (e.g$.$ Elmegreen
2001) would result in ubiquitous and rather continuous star formation,
with the UV output being dominated by young, luminous regions that
also emit in H$\alpha$.

Therefore, although the difference between the H$\alpha$ and UV radial
profiles might be partly due to a lack of massive stars in the
clusters of the XUV disk or to a higher porosity of the ISM to the
ionizing photons (Meurer et al$.$ 2004), especially in its outermost
regions, the use of azimuthal averages computed over spatial scales
with dynamical time-scales that are much longer than the evolutionary
time-scale of the emission under study (like is the case of H$\alpha$)
artificially leads to a dimming in the outer parts of the
corresponding surface-brightness profile.

\subsection{On the origin of the XUV emission}

As we mentioned above, NGC~4625 is not the only galaxy in which GALEX
has found extended, UV-bright disks with no obvious or very faint
optical counterparts. Other remarkable objects are M~83, NGC~5055 and
NGC~2841. Early results indicate that the presence of a large H{\sc i}
disk is a necessary but not sufficient condition for the XUV emission
to be present (Thilker et al$.$ 2005b, in prep.). Objects with
characteristics similar to those of NGC~4625, i.e$.$ galaxies in
interacting and/or paired SBm systems showing extended neutral-gas
disks (like NGC~3664 or NGC~4027) might be also good candidates to
show XUV emission.

In the case of NGC~4625 the star formation in the out H{\sc i} disk is
probably due to gravitational instability associated with the
interaction with NGC~4618 and possibly also with NGC~4625A. In this
sense, Bush \& Wilcots (2004) argue that NGC~4618 and NGC~4625 have
only had one close passage and that the current interaction has been
ongoing for $\sim$0.5\,Gyr. This, we note, is of the same order as the
time-scale of the UV emission. Since NGC~4618 does not show XUV
emission despite having a relatively extended H{\sc i} envelope
itself, there must be additional factors regulating the formation of
XUV disks.

In this context, we note that NGC~4625 shows a very regular velocity
field with differential rotation, while the kinematics of NGC~4618 are
highly disturbed, with some features characteristic of a strongly
warped disk. It is therefore possible that a relatively undisturbed
gas disk is required for the stars responsible for the XUV emission to
form. In the case of NGC~4625 the stability of the H{\sc i} disk is
thought to be a consequence of its small
$M_{\mathrm{disk}}$/$M_{\mathrm{halo}}$ mass ratio (Dubinski, Mihos,
\& Hernquist 1996).
\\

In summary, NGC~4625 provides a unique opportunity to study the
physics of star formation in the early stages of the formation of
disks in spiral galaxies. It is worth noting that although the star
formation in the XUV disk of NGC~4625 is believed to be due to the
presence of a nearby companion (NGC~4618 and/or NGC~4625A), this
system should not necessarily be viewed as an anomaly, since frequent
interactions are thought to be also a fundamental ingredient in the
initial growth of disks of spiral galaxies at high redshift.

\acknowledgments

GALEX is a NASA Small Explorer launched in April 2003. We gratefully
acknowledge NASA's support for construction, operation, and scientific
analysis for the GALEX mission. We thank Judith Cohen for kindly
providing her H$\alpha$ filter for COSMIC. RCK is partially financed
by NASA grant NAG5-8426.

\begin{figure}
\epsscale{0.9}
\plotone{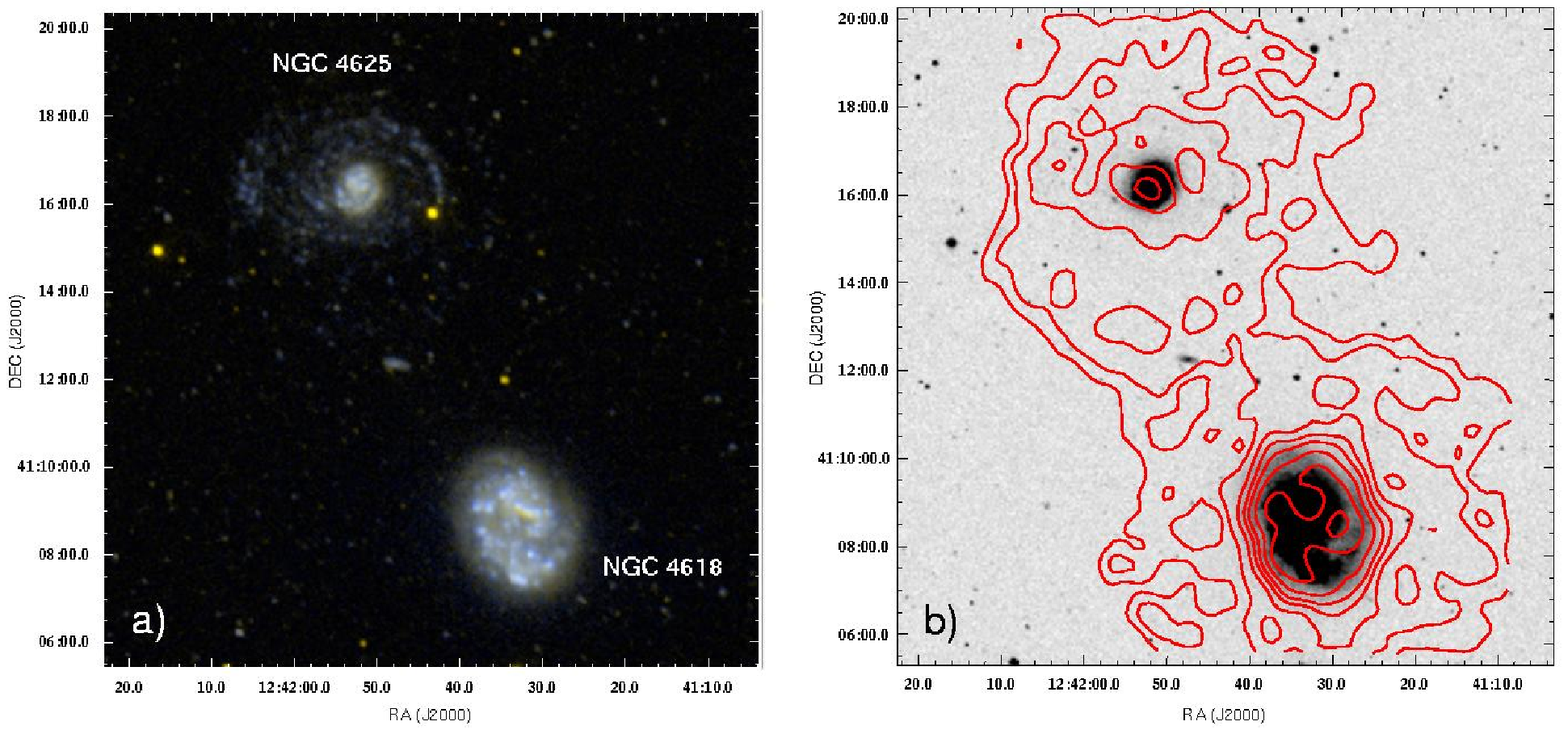}
\caption{{\bf a)} GALEX false-color RGB composite image (R=NUV; G=0.2$\times$FUV$+$0.8$\times$NUV; B=FUV) of the NGC~4618/NGC~4625 system with an arcsinh stretch (Lupton et al$.$ 2004). {\bf b)} POSS2 blue-plate DSS image of the same region with the 21\,cm H{\sc i} contours overimposed (1 2 4 6 8 11 15 20 $\times$10$^{20}$\,cm$^{-2}$). The H{\sc i} contours were obtained from our 30\,arcsec-resolution convolved map. The 3\,$\sigma$ level of this H{\sc i} map is at 1.5$\times$10$^{20}$\,cm$^{-2}$. The morphology of the XUV disk is remarkably similar to that of the H{\sc i} disk.\label{fig1}}
\end{figure}
\begin{figure}
\epsscale{1.0}
\epsscale{0.9}
\plotone{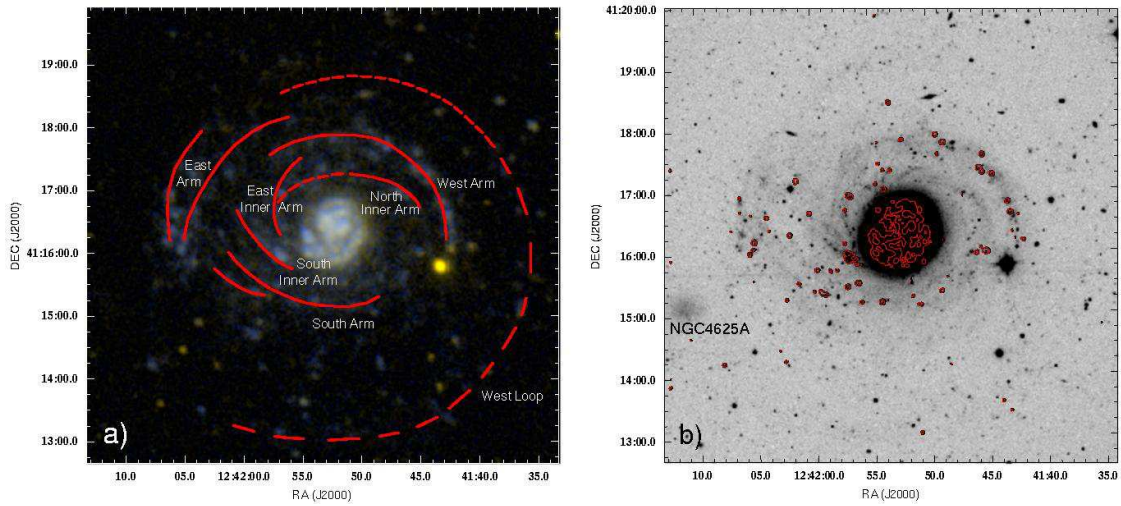}
\caption{{\bf a)} GALEX false-color image of NGC~4625 (color scheme as in Figure~1). A sketch of the spiral morphology of the XUV disk is also shown. Long-dashed lines show the position of the West Loop. Short-dashed lines represent tentative extensions of some of the structures identified in the UV images. {\bf b)} Deep ground-based $B$-band image of the same region. We have labelled the newly discovered companion of NGC~4625 as NGC~4625A. Contours of the continuum-subtracted H$\alpha$ emission are overplotted.\label{fig2}}
\end{figure}

\begin{figure}
\epsscale{0.8}
\plotone{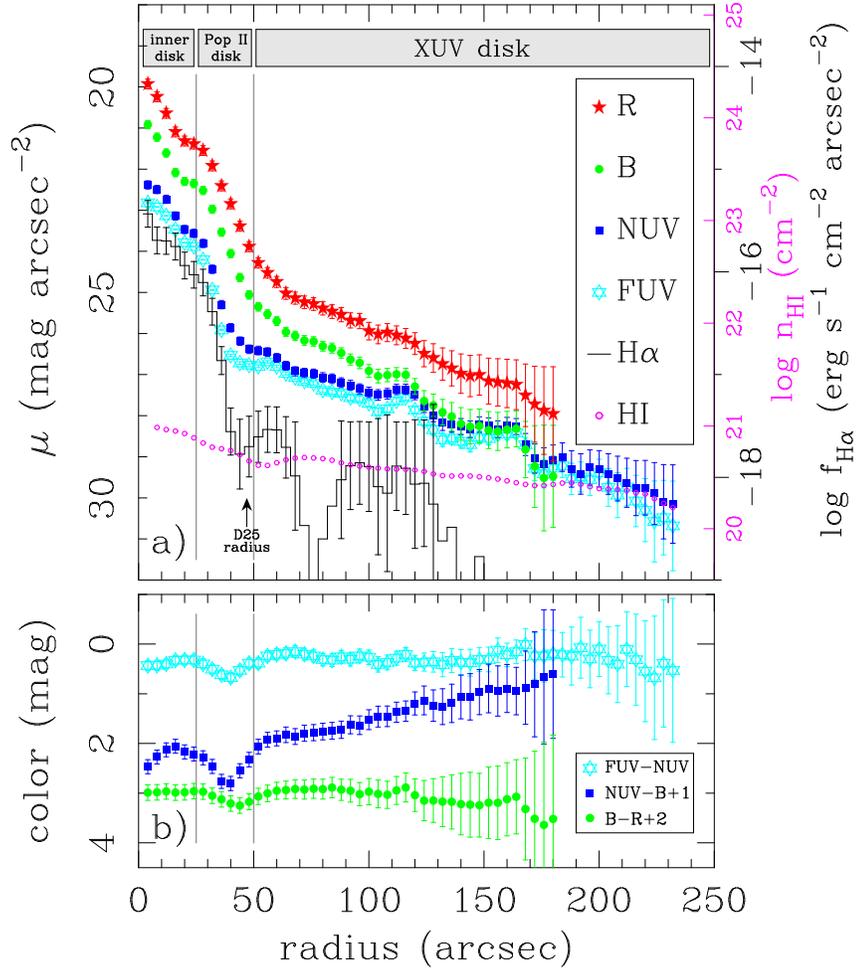}
\caption{{\bf a)} Surface-brightness profiles of NGC~4625. Error bars do not include calibration uncertainties. Error bars in the H$\alpha$ profile larger than 1\,dex have been removed for the sake of clarity. The UV emission clearly extends beyond four times the D25 radius (47\,arcsec). UV magnitudes are in AB scale while optical photometry is in the Johnson/Cousins system. {\bf b)} Observed (FUV$-$NUV), (NUV$-$$B$), and ($B-R$) color profiles. The (FUV$-$NUV) and ($B-R$) color profiles are remarkably flat except at the position where the (red) Population II component dominates the emission. On the other hand, the (NUV$-$$B$) profile gets systematically bluer towards the outer parts of the XUV disk. Note that the (NUV$-$$B$) and ($B-R$) color profiles are off-set by 1 and 2\,mag respectively for clarity of presentation on a single plot.\label{fig3}}
\end{figure}
\begin{figure}
\epsscale{0.8}
\plotone{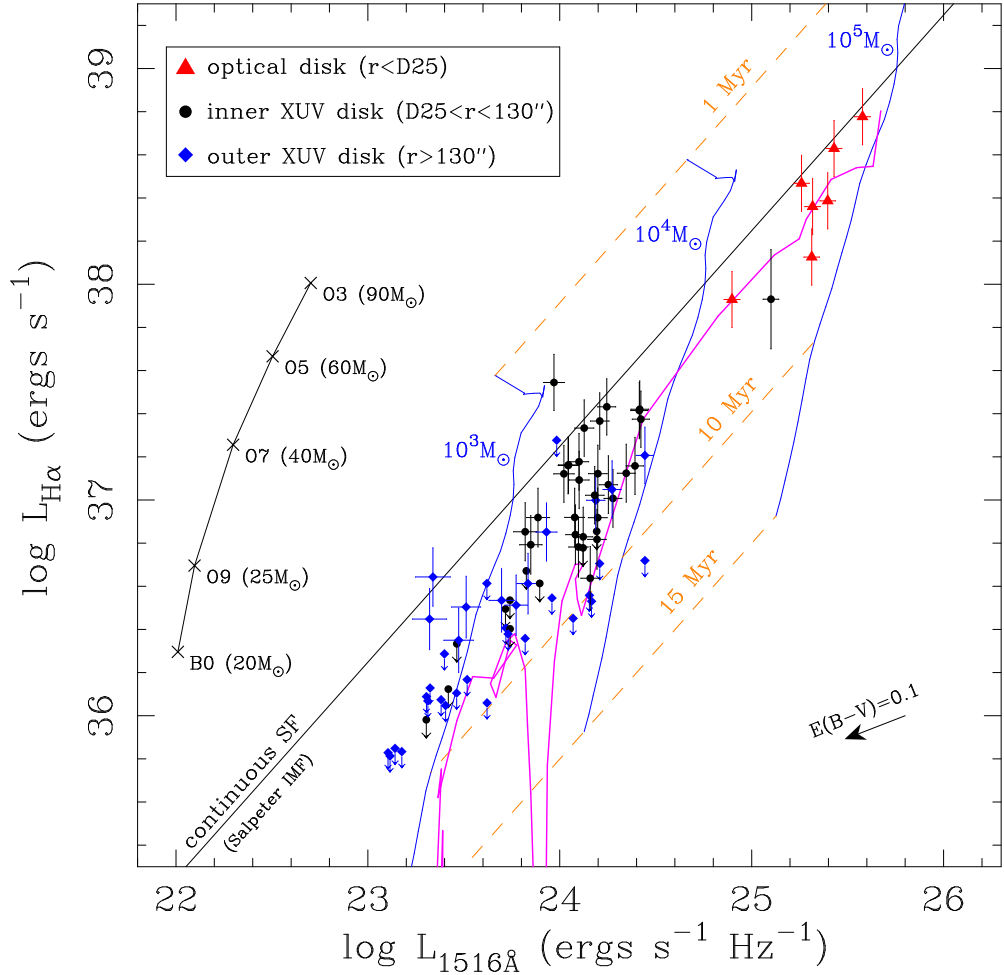}
\caption{H$\alpha$ vs$.$ FUV luminosity of FUV-selected complexes inside D25 (trianges), in the inner XUV disk (dots) and the outer XUV disk (diamonds). We also show the model predictions for the luminosities of single massive stars (Vacca, Garmany, \& Shull 1996; Sternberg, Hoffmann, \& Pauldrach 2003) and instantaneous bursts with different ages and masses for a Salpeter IMF with mass range 0.1-100\,M$_{\odot}$ (Bruzual \& Charlot 2003). Model curves are plotted without any reddening. The thick solid magenta line represents the expected luminosity inside an average aperture of 7\,arcsec in radius and a surface-brightness given by the azimuthally-averaged profiles shown in Figure~3. A reddening vector of E($B-V$)=0.1\,mag is also shown in the lower right corner.\label{fig4}}
\end{figure}

\end{document}